\documentclass[journal]{IEEEtran}
\usepackage{amsmath,amsfonts}

\usepackage{array}
\usepackage[caption=false,font=normalsize,labelfont=sf,textfont=sf]{subfig}
\usepackage{textcomp}
\usepackage{stfloats}
\usepackage{url}
\usepackage{verbatim}
\usepackage{graphicx}
\usepackage{cite}
\usepackage{multirow}
\usepackage{booktabs}
\usepackage{graphicx} 
\usepackage[utf8]{inputenc}
\usepackage[ruled,vlined]{algorithm2e}

\usepackage{graphicx}

\SetKwInOut{Input}{Input}
\SetKwInOut{Output}{Output}

\SetAlgoLined

\usepackage{soul} 
\usepackage{xcolor} 
\usepackage[most]{tcolorbox}
\usepackage{float}  
\usepackage{orcidlink}

\sethlcolor{red} 

\begin{document}

\title{Bridging Bug Localization and Issue Fixing: A Hierarchical Localization Framework Leveraging Large Language Models}

\author{Jianming Chang \orcidlink{0000-0002-0052-6198}, Xin Zhou \orcidlink{0000-0002-4558-0622}, Lulu Wang \orcidlink{0000-0002-8575-4172}, David Lo \orcidlink{0000-0002-4367-7201}, \IEEEmembership{Fellow,~IEEE}, Bixin Li \orcidlink{0000-0001-9916-4790}
       
\thanks{This work is supported by the National Natural Science Foundation of China under Grant No. 61872078. (Corresponding authors: Xin Zhou; Lulu Wang.)}
\thanks{Jianming Chang, Lulu Wang, and Bixin Li are with Southeast University, China 211189 (e-mail: \{jianmingchang, wanglulu, bx.li\}@seu.edu.cn).}
\thanks{Xin Zhou, and David Lo are with Singapore Management University, Singapore 188065 (e-mail: davidlo@smu.edu.sg; xinzhou.2020@phdcs.smu.edu.sg).}
}

\maketitle

\begin{abstract}
Automated issue fixing is a critical task in software debugging and has recently garnered significant attention from academia and industry. However, existing fixing techniques predominantly focus on the repair phase, often overlooking the importance of improving the preceding bug localization phase. As a foundational step in issue fixing, bug localization plays a pivotal role in determining the overall effectiveness of the entire process.

To enhance the precision of issue fixing by accurately identifying bug locations in large-scale projects, this paper presents BugCerberus, the first hierarchical bug localization framework powered by three customized large language models. First, BugCerberus analyzes intermediate representations of bug-related programs at file, function, and statement levels and extracts bug-related contextual information from the representations. Second, BugCerberus designs three customized LLMs at each level using bug reports and contexts to learn the patterns of bugs. Finally, BugCerberus hierarchically searches for bug-related code elements through well-tuned models to localize bugs at three levels. With BugCerberus, we further investigate the impact of bug localization on the issue fixing.

We evaluate BugCerberus on the widely-used benchmark SWE-bench-lite. The experimental results demonstrate that BugCerberus outperforms all baselines. Specifically, at the fine-grained statement level, BugCerberus surpasses the state-of-the-art in Top-N (N=1, 3, 5, 10) by 16.5\%, 5.4\%, 10.2\%, and 23.1\%, respectively. Moreover, in the issue fixing experiments, BugCerberus improves the fix rate of the existing issue fixing approach Agentless by 17.4\% compared to the best baseline, highlighting the significant impact of enhanced bug localization on automated issue fixing.
\end{abstract}


\section{Introduction}

Automated GitHub issue fixing has recently garnered significant attention from researchers and developers for its potential to alleviate the inefficiencies of manual software debugging and repair~\cite{jimenez2024swebench, agentless, Zhang2024}. However, issue fixing remains a challenging task due to the need for in-depth analysis of extensive codebases and intricate functionalities. Various automated approaches have been proposed in recent years, which mainly focus on enhancing the accuracy of the issue fixing.

Most existing automated issue-fixing approaches divide the task into two major phases: 1) bug localization and 2) issue resolution. Bug localization aims to identify suspicious parts of the entire code repository responsible for the issue. The subsequent issue resolution phase seeks to automatically generate fixing patches using the bug location, issue description, and source code. 
For instance, RAG~\cite{Carlos2024} and CodeR~\cite{chen2024coder} utilize BM25 to compute the textual similarity between the problem description and code files, localize bugs at the file level, and attempt repairs within the identified files using the repair agent.

Bug localization is a critical prerequisite for effective issue fixing. However, most existing approaches primarily focus on improving techniques for the repair phase while relying on basic methods for bug localization. As a result, the potential of bug localization remains under-explored. We identify three key \textbf{limitations} in the current bug localization methods:

$\bullet$ \textbf{Limited code context.} Existing methods, such as RAG~\cite{Carlos2024} and CodeR~\cite{chen2024coder}, rely on BM25 to calculate the similarity between code and bug descriptions. However, in real-world projects, code semantics extend beyond the code itself; the surrounding context plays an equally critical role~\cite{Li2020}. As a result, enhancing the ability to recognize code semantics within its broader context is essential for improving bug localization and, ultimately, making issue fixing more effective.

$\bullet$ \textbf{Missing hierarchical and customized code understanding.} Although many current issue fixing methods~\cite{agentless, Zhang2024} adopt a hierarchical approach (file-function-statement) to achieve statement-level bug localization, they often apply the same code analysis techniques across all three levels. However, the semantic content and dependencies in code differ significantly at each level. For instance, function-level code relies on function calls, while statement-level code depends on data and control dependencies between statements. Furthermore, the challenges of learning bugs vary across levels. At the coarse-grained file level, models face the limitation of processing large-scale code inputs, which can slow down computation and increase the cost of the localization process. Addressing this requires strategies to extract a reasonable length of context. At the fine-grained function and statement levels, where the semantics may be limited, it is essential to extract inter-function and inter-statement dependencies to enrich the code context. Therefore, customized code-learning techniques should be developed to accommodate the varying granularities of bug localization.

$\bullet$ \textbf{Low accuracy in fine-grained bug localization.} The accuracy of bug localization directly determines the upper bound of effectiveness for localization-based issue fixing methods. Although existing approaches achieve satisfactory results at coarse-grained levels, such as the file level, their performance at finer granularities, like the statement level, still leaves significant room for improvement. Fixing experience suggests that finer localization granularity often leads to more effective issue resolution~\cite{Meng2022}.

\begin{figure*}[!t]
	\centering
	\includegraphics[width=1\textwidth]{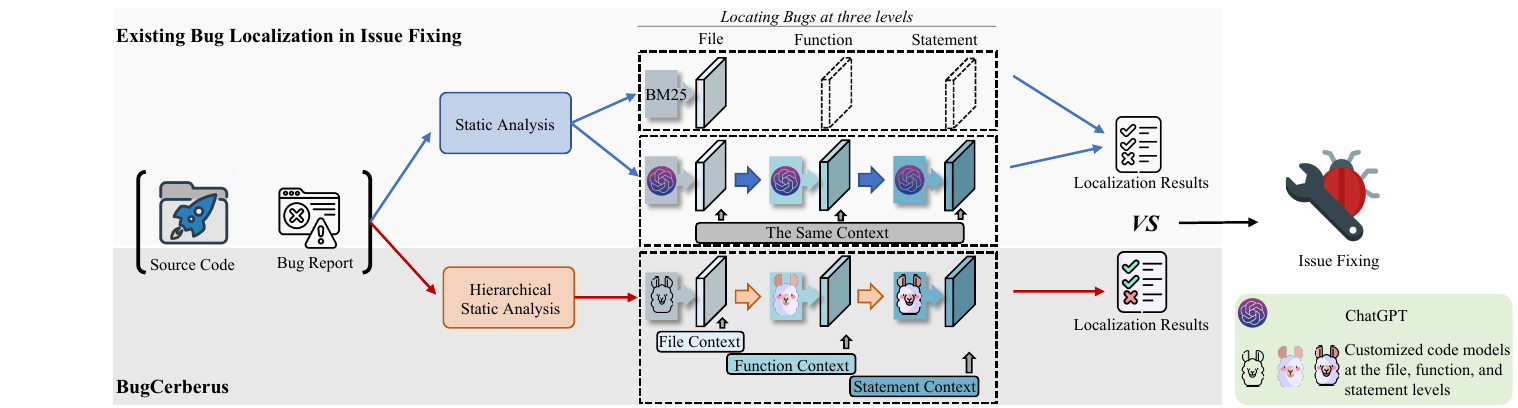}
	\caption{Difference between BugCerberus and Existing Bug Localization Methods in Issue Solving. Unlike existing methods, BugCerberus utilizes customized contexts at the file, function, and statement levels through hierarchical static analysis. Additionally, BugCerberus incorporates three specialized code models tailored for each of these levels.}
	\label{IntroExample}
\end{figure*}

\textbf{Our work}. To address these limitations, we propose a hierarchical learning-based bug localization method, as illustrated in Figure \ref{IntroExample}. Compared with existing bug localization methods in issue fixing, BugCerberus is capable of locating bugs at finer granularities and capturing corresponding code semantics at the file, function, and statement levels through the hierarchical static analysis and three customized code models. While dynamic analysis can provide valuable context by executing test cases, it may not always be feasible when bugs occur due to unavailable or failing test cases. Therefore, BugCerberus employs static analysis to extract contextual information. At each level, BugCerberus constructs relationships between code elements and extracts their semantics based on these relationships. It then learns the relevance between buggy code and bug descriptions. By ranking these relevance scores, BugCerberus achieves accurate bug localization across all levels. Existing research has demonstrated the effectiveness of the Llama model in various software engineering tasks~\cite{Xin2024, Guan2024, Tu2024, Sun2024, Hou2024}. To further enhance the understanding of buggy code, our method fine-tunes Llama 3.1 to learn semantic connections between buggy code and problem descriptions.

To evaluate the effectiveness of BugCerberus, we conducted a comprehensive experiment on the dataset SWE-bench-lite~\cite{Carlos2024}. We compared our approach with Agentless \cite{agentless} and FBL-BERT \cite{Ciborowska2022}. Agentless is the best open-source issue-fixing approach, which contains bug localization and repair phases, while FBL-BERT is the state-of-the-art method for bug localization. The experimental results show that our approach achieves the best results on all metrics at the three levels. Specifically, for the localization at the finest-grained statement level, BugCerberus improves over the best baseline by 16.5\%, 5.4\%, 10.2\%,  23.1\% in Top-N (N = 1, 3, 5, 10), 57.3\% in MRR, and 55.3\% in MAP correspondingly. BugCerberus is designed to improve the bug localization part of the issue fixing approach. To assess its impact on issue fixing, we choose open-source Agentless \cite{agentless}, as we do not have access to many agents released by companies. The experiment shows that BugCerberus boosts the performance of Agentless by 17.4\% with our more accurate bug localization results, demonstrating the importance of bug localization in issue fixing. Additionally, we analyze the individual contributions of each component and find that every component contributes to the overall performance of BugCerberus.

In summary, our contributions are as follows:
\begin{itemize}
	\item  
	
	To the best of our knowledge, BugCerberus is the first LLM-based framework mainly designed to enhance the bug localization phase of current issue fixing techniques. BugCerberus analyzes distinct semantic representations for code at three granularity levels and designs three customized LLMs at each level to identify the suspicious code responsible for the bug.
	
	\item  We investigate the impact of bug localization on issue fixing. We compare BugCerberus with state-of-the-art bug localization approaches, including those applied in recent issue fixing techniques. Furthermore, we explore how the bug localization approaches contribute to the effectiveness of the issue fixing.
	
	\item  We release the context dataset across the three granularity levels. We construct a dataset that captures the context of buggy-related code elements at the file, function, and statement levels. This dataset is publicly available to support future research on LLM-based bug localization approaches and to facilitate the study of bug characteristics across different granularity levels.
	
\end{itemize}

\section{Background}

This section provides the background related to BugCerberus, including the instruction learning, program dependency graph, and GitHub issue fixing dataset.

\subsection{Instruction Learning}
Instruction learning~\cite{wei2022}, also referred to as instruction fine-tuning, involves training models to comprehend natural language instructions and execute corresponding tasks. This process typically combines supervised learning and reinforcement learning~\cite{Yin2024}. In supervised learning~\cite{Sanh2022}, the model is fine-tuned on a diverse set of example instructions paired with their corresponding outputs. In reinforcement learning~\cite{Ouyang2024}, the LLM adjusts its parameters based on feedback from the environment and associated reward signals. Task instructions often comprise multiple components; in this study, we adopt a prior instruction design framework~\cite{Sanh2022, ZhangQuanjun2023}, incorporating instructions with input-output examples. Utilizing the buggy codes from the SWE-bench dataset, we annotate bug-related and unrelated code elements and apply a supervised learning approach to guide the fine-tuning of the LLM.

\begin{figure*}[!t]
	\centering
	\includegraphics[width=1\textwidth]{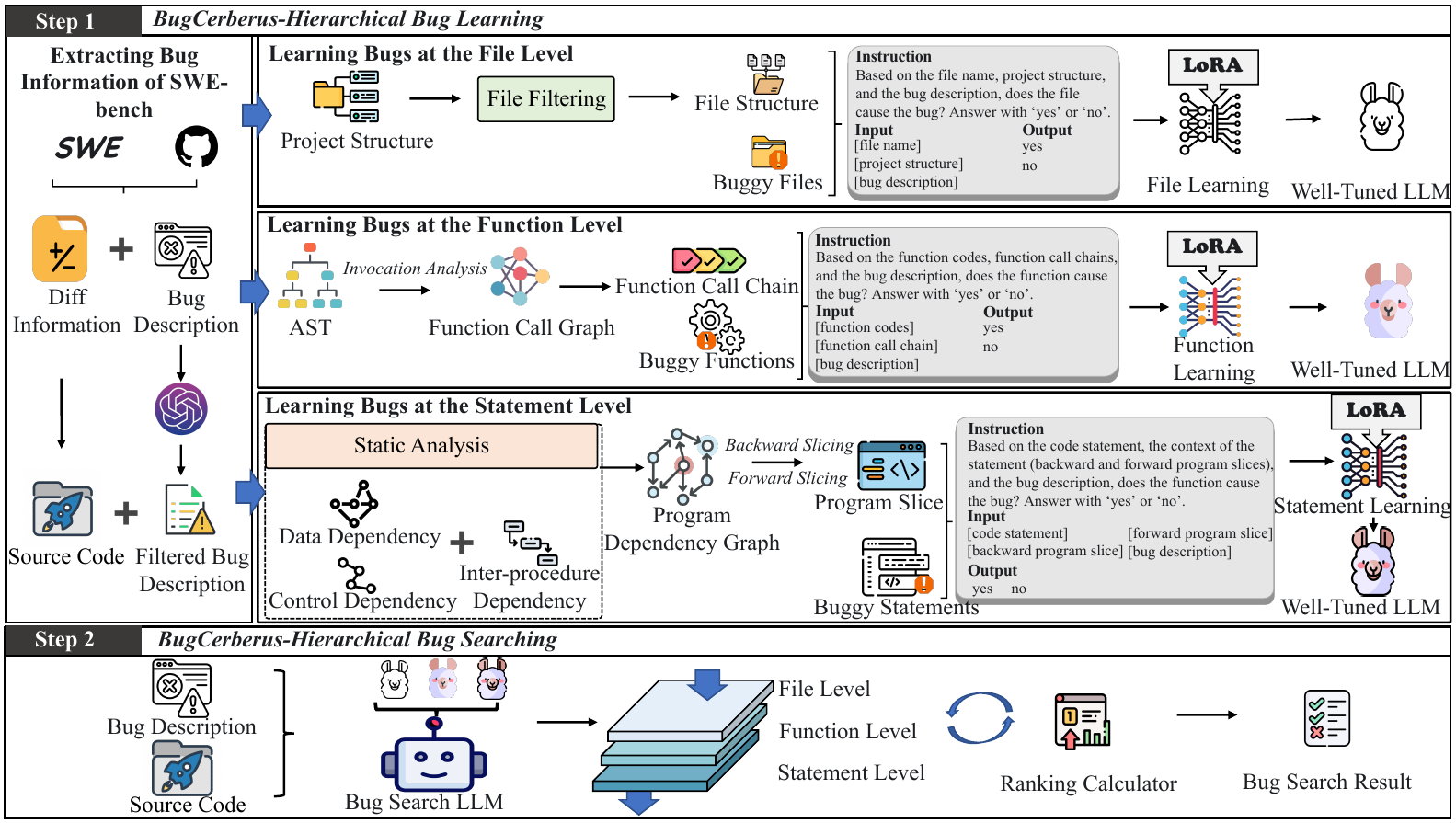}
	\caption{Overview of Our Approach}
	\label{workframe}
	\vspace{-0.3cm}
\end{figure*}

\subsection{Program Dependency Graph}

A program dependency graph (PDG)~\cite{Jeanne1987, Susan1992} is an intermediate representation of a program, structured as a directed graph that captures both data dependencies and control dependencies within a procedure. The computation of control dependencies is based on the program’s control flow~\cite{Jeanne1987}. Specifically, node \emph{i} is control dependent on node \emph{j} if the following two conditions are met: (1) there exists a directed path from \emph{j} to \emph{i} in the control flow graph, and \emph{i} is not post-dominated by \emph{j}; (2) \emph{i} is post-dominated by at least one successor of \emph{j}. For data dependency extraction, the focus lies on the definition and use of variables~\cite{Jeanne1987} : statement \emph{a} is data dependent on statement \emph{b} if \emph{a} uses a variable that is defined in the previous statement \emph{b}. Furthermore, we extend the PDG by incorporating invocation relationships between statements and functions, enabling it to support inter-procedural dependency analysis~\cite{Susan1990}. In this paper, the extended PDG is utilized for program slicing to extract the program's contextual information effectively.

\subsection{GitHub Issue Fixing Dataset}

GitHub is a social coding platform hosting over 57 million repositories~\cite{Panichella2021}, providing an integrated issue tracker. However, issues often are noisy and poorly documented, making automated issue-fixing challenging. To address this, Jimenez et al.~\cite{jimenez2024swebench} introduced SWE-bench, a high-quality benchmark of GitHub issues collected from 12 popular open-source Python repositories. Each instance in SWE-bench includes (1) an issue description, which is a natural language statement describing the issue; (2) a base commit, representing the commit ID on which the original pull request was applied; and (3) a fixing patch, extracted from the code changes in the original pull request that resolved the issue. In this paper, BugCerberus leverages the issue information to learn the semantics of issues and improve bug localization.

\section{Our Approach}

\textbf{Framework.} We introduce BugCerberus, an effective LLM-powered framework for multi-level bug localization, encompassing file, method, and statement levels. Figure \ref{workframe} shows the overall workflow of BugCerberus. Given the project containing the bug and the available bug report, BugCerberus produces a ranking list of the top k most suspicious files, functions, and statements responsible for the bug. BugCerberus contains two main steps: 

\begin{itemize}
	\item \textbf{Step 1: Hierarchical Bug Learning.} In this step, BugCerberus analyzes the static information of the buggy codes at the three levels and fine-tunes three LLMs to learn the buggy code patterns specific to each level.
	
	\item \textbf{Step 2: Hierarchical Bug Searching.} In this step, the fine-tuned models are employed in a hierarchical search process, identifying the bug progressively at the file, method, and statement levels.
\end{itemize}

\subsection{Hierarchical Bug Learning}
This section details the extraction of bug information of SWE-bench, the retrieval of bug-related context at the file, function, and statement levels, and the learning process of three large-scale models for bug localization.

\subsubsection{Extracting Bug Information of SWE-bench}

SWE-bench collects the code segments that must be modified to fix an issue, treating these segments as buggy code~\cite{Meng2022, YangHaoran2024, Michael2020}. Using the commit ID before the fix, we retrieved the bug-containing project source code. This code serves as the data for subsequent bug analysis.

\begin{figure*}[!t]
	\centering
	\includegraphics[width=1\textwidth]{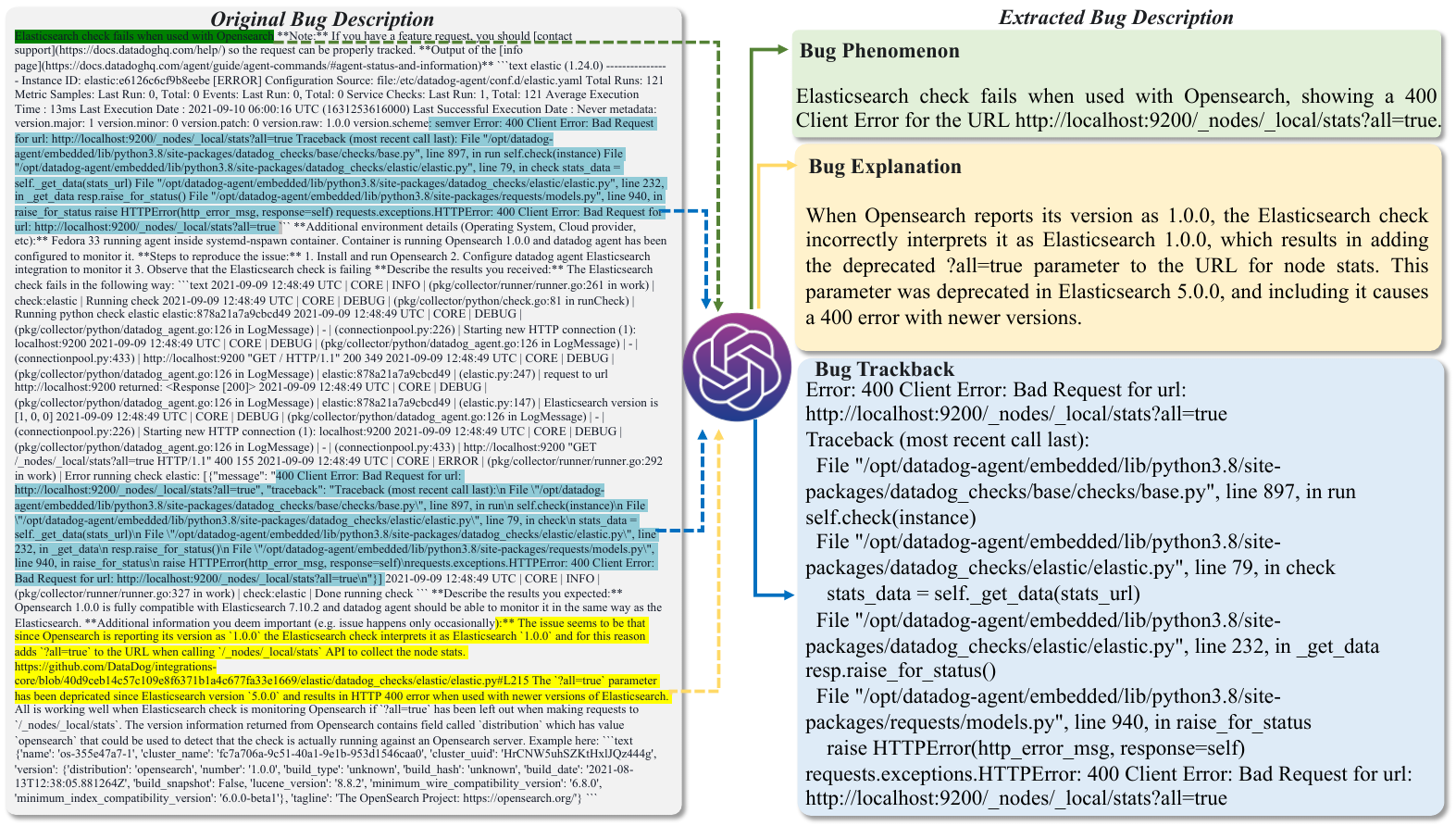}
	\caption{Example of Bug Report Information Extraction}
	\label{extractedProblem}
	\vspace{-0.3cm}
\end{figure*}

\begin{figure}[!t]
	\centering
	\includegraphics[width=0.4\textwidth]{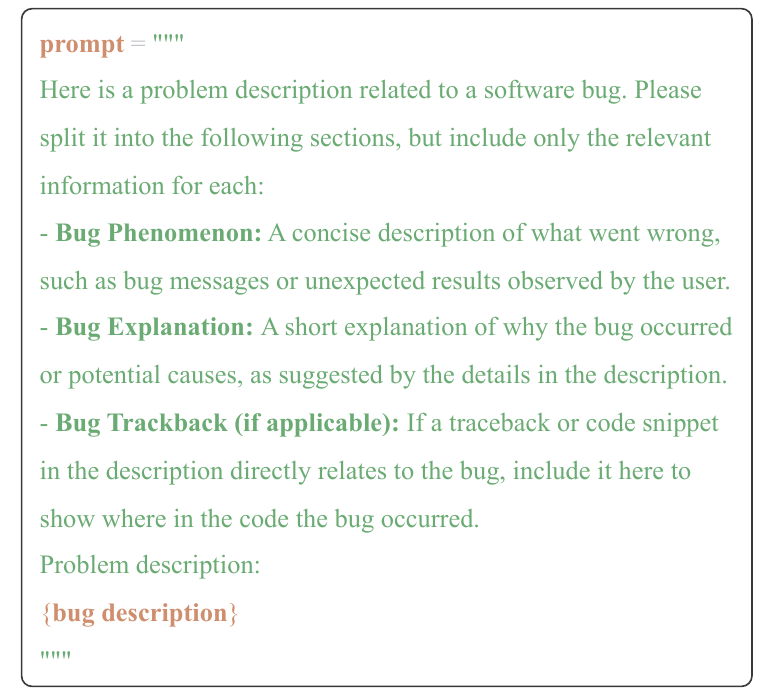}
	\vspace{-0.5cm}
	\caption{Prompt for Information Extraction}
	\label{extractionprompt}
	\vspace{-0.2cm}
\end{figure}

In our analysis of bug descriptions, we discovered that many bug descriptions in the SWE-bench dataset are quite complex. For the instance of Figure \ref{extractedProblem}, the bug description with the instance ID 	
\emph{DataDog\_integrations-core-10093} contains 6,437 characters. This complexity can hinder the model's ability to learn the relevant features of the bugs~\cite{ZHANG2015, Chris2018}. Previous studies have effectively demonstrated ChatGPT's ability to summarize bug descriptions~\cite{Leong2024, Bo2024, Zhang2023}. In this paper, we leverage ChatGPT~\cite{openai2023gpt4} to extract key information from bug descriptions concisely. Specifically, ChatGPT (the GPT-4 2024-05-13 model) is used to identify and summarize important information about the bug, including the observed phenomenon, the cause for the bug, and the bug traceback. We extracted these three components using the prompt in Figure \ref{extractionprompt} with default ChatGPT hyperparameters. 

Taking Figure \ref{extractedProblem} as an example, ChatGPT identifies the bug phenomenon that the \emph{Elasticsearch check} fails when used with \emph{OpenSearch} from the complex bug description and extracts the reasons behind this bug. Simultaneously, ChatGPT gathers information about the fault traceback of the four executed code files using the provided prompt. These extracted details are then used for subsequent model fine-tuning, replacing the need to use the entire bug description.

\subsubsection{Bug learning at the three levels}
In this subsection, we explain how to conduct customized bug learning at the file, function, and statement levels.

\begin{figure}[!t]
	\centering
	\includegraphics[width=0.5\textwidth]{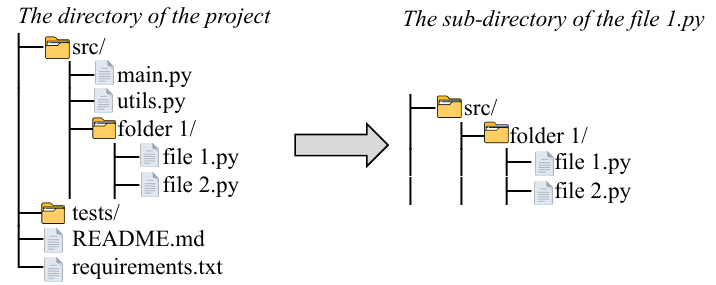}
	\vspace{-0.5cm}
	\caption{Example of a Subdirectory}
	\label{directoryExample}
	\vspace{-0.3cm}
\end{figure}

\textbf{File Level.}
Analyzing the entire repository at once is too lengthy for an LLM to handle directly and would significantly slow down the computation. Therefore, we simplify the input by restricting the analysis to a smaller directory scope. For a given project, we first construct its directory structure and then extract the sub-directory structure encompassing the target file. As illustrated in Figure \ref{directoryExample}, the constructed structure arranges files and folders vertically at the same directory level, while sub-directories are represented with indentation. This sub-directory structure preserves all relevant folders along the file path.

The sub-directory structure serves as an input to the model, providing information about the neighboring files and the initial project structure, and aligns with the convention that most files at the same directory level serve similar functional requirements. Additionally, we retain only Python files within each directory, excluding test files or other formats like TXT, as the files are beyond the scope of our bug localization task. Although relying solely on the directory structure may not be fully accurate, we adopt it as a coarse-grained filtering step, as the subsequent fine-grained localization will enable BugCerberus to locate the bug more precisely. Therefore, extremely high accuracy is not required at this initial coarse-grained stage.

\begin{figure}[!t]
	\centering
	\includegraphics[width=0.4\textwidth]{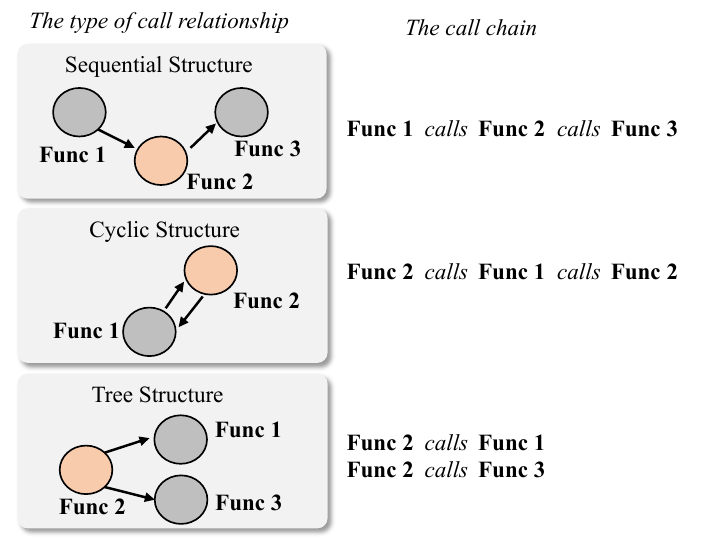}
	\vspace{-0.5cm}
	\caption{Call Chain Analysis}
	\label{methodInfo}
	\vspace{-0.5cm}
\end{figure}

\textbf{Function Level.}
We use the function call graph as the representation to capture function-level semantics. The function call graph captures the invocation relationships between functions, providing a means to identify functions related to the buggy function~\cite{Le2018}. Furthermore, the function call chains extracted from the graph serve as valuable contextual information, aiding the bug localization process in understanding the semantics of the buggy function~\cite{Beszédes2020, ChenXin2024}. We construct the Abstract Syntax Tree (AST) of the program and extract function call relationships to build a function call graph. Using this graph, we iteratively compute the forward and backward call chains of the target function, representing its partial content and its impact scope, respectively. The extracted call chains are connected by the keyword `call,' forming textual representations like `func1 calls func2 calls func3' as textual input required by the model. For complex dependency paths in the call graph, we retain the complete dependency information. As illustrated in Figure \ref{methodInfo}, for cyclic structures, the start and end points of the call chain coincide, indicating a cyclic dependency path. For tree structures, we enumerate all reachable paths. Both the function content and its call chains are used as inputs to the model to capture the correlation between the function and the problem description.

\textbf{Statement Level.}
Statement-level bug localization is limited by the minimal semantic information provided by the previously localized function. However, contextual information from other related functions can aid in identifying fine-grained, line-level locations. To address this limitation, we leverage program slicing~\cite{Weiser1984} techniques to capture valuable dependencies that provide a comprehensive context for each statement. By leveraging program slices, LLMs can better understand the semantics of buggy statements~\cite{Cao2022,Jiang2024,WangChe2024}. We utilized Joern~\cite{joern2025} to construct a program dependency graph that captures the dependencies within a single procedure. Subsequently, we extended the graph to include inter-procedural dependencies by analyzing invocation relationships. The resulting graph integrates intra-procedural dependencies, such as data and control dependencies, alongside inter-procedural dependencies, including function calls. We excluded statements without substantive content, such as comments, blank lines, and purely syntactic symbols.

With the constructed program dependency graph, we obtain the context of the statement by performing forward program slicing and backward program slicing using the statement as the slice criterion. The forward program slice reflects the impact scope of the statement, the backward program slice contains the scope on which the statement depends, and both the forward and backward slices describe the context of the statement. Besides, the program slice can provide the location of the statement, like the reachable function, and makes the model to well distinguish the statement. Program slicing is a reachable analysis on the constructed graph, Algorithm \ref{slicingAl} describes the backward slicing on the intra and inter-procedural dependencies, while the forward slicing is revised.

The program slicing algorithm is based on a two-phase approach~\cite{Susan1990} and comprises four steps. For instance, in the case of backward slicing, the process begins with Step 1, where the algorithm identifies the node \emph{N$_c$} corresponding to the slicing criterion in the dependency graph. In Step 2, the algorithm performs a backward traversal starting from \emph{N$_c$} to determine the intra-procedural reachable node \emph{N$_1$}. During this traversal, the scope is restricted to data and control dependencies to prevent premature traversal into the callee procedure. Next, in Step 3, all backward-reachable nodes \emph{N$_2$} are identified starting from \emph{N$_1$}. A dependency restriction `\emph{Return}' is applied to avoid redundant entries into the caller process, thereby eliminating unnecessary iterations. Finally, in Step 4, the location attributes of the reachable nodes, including their corresponding files and lines of code, are utilized to map back to the relevant code content, ultimately generating the backward slices associated with the slicing criterion.

\begin{figure*}[!t]
	\centering
	\includegraphics[width=\textwidth]{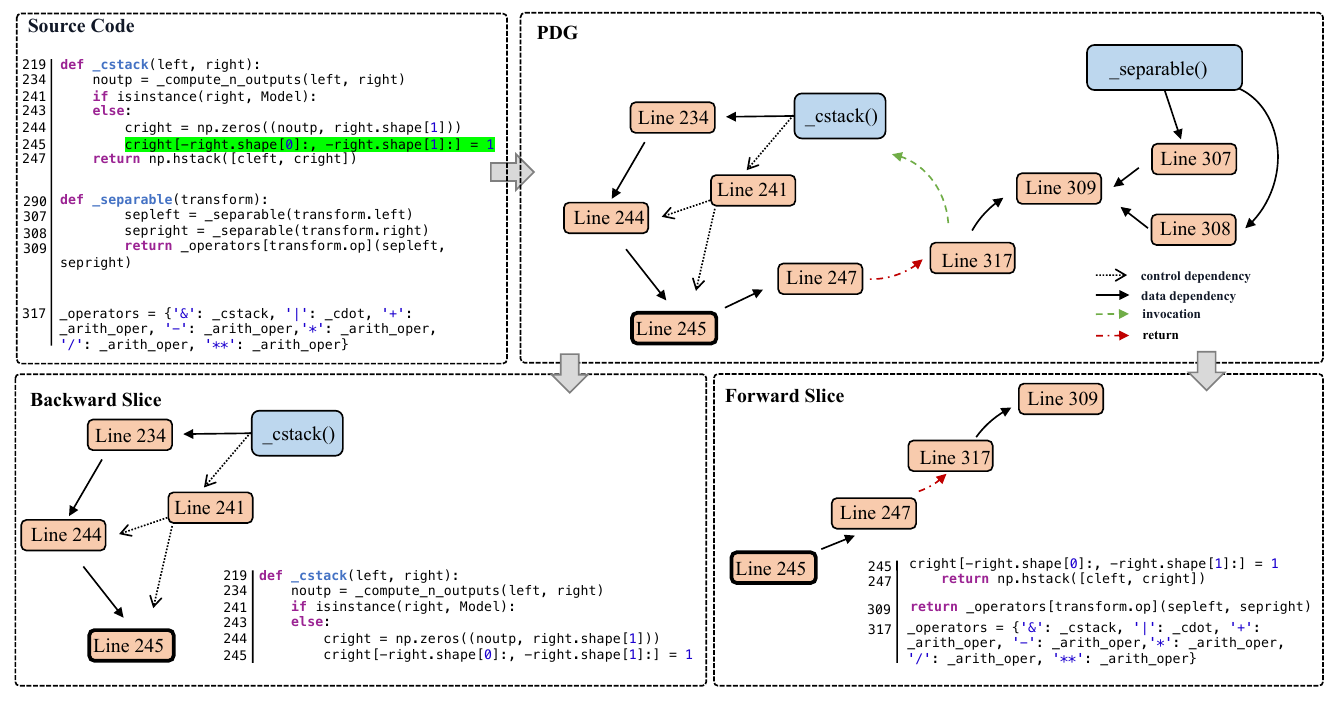}
	\vspace{-0.5cm}
	\caption{PDG Example and Program Slice for Line 245}
	\label{SliceExample}
	\vspace{-0.5cm}
\end{figure*}

\begin{algorithm}[!t]
	\caption{Program Slicing Algorithm}
	\KwIn{CPG, \emph{slicing criterion}: SC $\langle$file,line$\rangle$}
	\KwOut{backward slice: BS, forward slice: FS}
	
	\BlankLine
	\textbf{Step 1: Localizing criterion nodes} \\
	Initialize N$_c$ = []\\
	\For{Node n in CPG}{\If{n.file==SC.file and n.line ==SC.line}{add n to N$_c$} }

	\BlankLine
	\textbf{Step2: Backward slicing} \\
	Match $N_1 \xrightarrow{e} N_c$ \textbf{where} $e \in$ \{data dependency, control dependency\}\\
	Match $N_2 \xrightarrow{e^*} N_1$ \textbf{where} $e \neq$ `Return'\\
	Add $N_1$, $N_2$ in BS
	
	\BlankLine
	\textbf{Step 3: Forward slicing} \\
	Match $N_c \xrightarrow{e} N_3$ \textbf{where} $e  \in$ \{data dependency, control dependency\}\\
	Match $N_3 \xrightarrow{e^*} N_4$ \textbf{where} $e \neq$ `Invocation'\\
	Add $N_3$, $N_4$ in FS
	
	\label{slicingAl}
	
\end{algorithm}

The forward slicing process follows a similar structure to that of backward slicing, with the primary difference being the direction of the dependency traversal, as illustrated in Algorithm \ref{slicingAl}.
Using the slicing algorithm, we can capture the context of the code statement. The code snippet in Figure \ref{SliceExample} is sourced from SWE-bench and corresponds to the instance ID astropy\_astropy-12907. Line 245, identified as the buggy code, is designated as the slicing criterion in the slicing algorithm. A program dependency graph is constructed for the code, and dependency analysis is performed to generate both backward and forward slicing results, using line 245 as the starting point. In this example, program slicing is employed to derive the comprehensive semantics of line 245, encompassing both the backward-dependent and forward-impact code. These semantics are then utilized to align the code context with bug descriptions, enabling the extraction of features related to the buggy code.

\subsubsection{Fine-tuning LLM for Bug Localization}

Based on the context of the buggy code extracted at three levels, as illustrated in Figure \ref{workframe}, we design specialized instructions for each model to focus on code semantics at the corresponding granularity. Each instruction primarily includes the buggy code itself, its surrounding context, and the problem description. Subsequently, we perform instruction fine-tuning on the LLM to enable it to learn the relationship between the buggy elements and the problem description.

To effectively learn from buggy elements, BugCerberus incorporates Flash Attention \cite{dao2022flashattention}, a highly efficient attention computation method. As outlined in Formula 1, Flash Attention divides the input sequence into multiple chunks using a sliding window chunking strategy. Each chunk independently computes the attention distribution with local normalization, ensuring numerical stability. The results from all chunks are then merged to produce the final global attention output. This approach significantly reduces memory consumption while maintaining high accuracy, making it particularly well-suited for processing long-sequence issue descriptions and source code.

\begin{equation}
	\text{Attention}(Q, K, V) = 
	\bigoplus_{i=1}^{m} 
	\left[ 
	\text{softmax}\left(\frac{Q_i K_i^\top}{\sqrt{d_k}} - M_i\right) V_i 
	\right]
\end{equation}

Within each chunk, Flash Attention computes the scaled dot-product attention, where $\frac{Q_i K_i^\top}{\sqrt{d_k}}$ represents the normalized similarity between query and key vectors. To ensure numerical stability, the maximum value $M$ is subtracted from the scores before applying the softmax function. The resulting attention weights are then used to compute a weighted sum of the value vector $V$, producing the chunk-wise output. The averaging fusion operation $\bigoplus$ ensures smooth transitions across overlapping regions of consecutive chunks. By averaging the attention outputs, this method avoids sharp discontinuities at boundaries while maintaining consistent attention distribution.

During training, we employ Low-Rank Adaptation (LoRA) to fine-tune the LLM efficiently. LoRA~\cite{hu2022lora} is a parameter-efficient fine-tuning technique designed to significantly reduce the number of trainable parameters. This approach has demonstrated its effectiveness in software debugging tasks~\cite{YangZhou2024, LiGuochang2024, Fan2024}. As shown in Formula 2, the core idea of LoRA involves freezing the pre-trained model's weights while introducing trainable low-rank matrices to capture task-specific updates.

\begin{equation}
	W' = W + \Delta W, \quad \Delta W = A \cdot B
\end{equation}
where $W$ is the frozen pretrained weight matrix,  $\Delta W $ is the task-specific weight adjustment, $A \in \mathbb{R}^{d \times r}$ is a trainable low-rank matrix, $B \in \mathbb{R}^{r \times k}$ is a trainable projection matrix, and $r \ll d$, $k$ is the rank of the low-rank decomposition.

\subsection{Hierarchical Bug Searching}

During the bug-searching phase, we utilize three models at the file, function, and statement levels to identify bug-related code elements. 
Specifically, similar to the training phase, we employ GPT4o to summarize bug-related descriptions from the original bug reports. To hierarchize and refine the scope of the bug search, we adopt a step-by-step approach, proceeding sequentially through files, functions, and statements. Given a project containing a bug, we first construct the project structure, extract the substructure of each file, and feed it into the file-level LLM to identify the Top-N bug-related files. Next, for these N files, we construct the call graphs of their functions and combine them with the function contents as input to the function-level LLM, generating a ranked list of the Top-N bug-related functions. Finally, for these N functions, we build associated program dependency graphs, incorporating both the functions themselves and their dependent code. We extract the forward and backward program slices for each statement within the functions and use them as input to the statement-level LLM to identify the Top-N bug-related statements.

For all three levels of the models, the input instructions are constructed using the same templates as in the training phase. At each level, we transform the representation of the ``yes" token in the output layer into a single value between 0 and 1 using an activation function. This transformed value is interpreted as the probability of the respective code element being buggy.

\section{Experiment Setup}

\subsection{Research Questions}
For evaluating BugCerberus, we consider the following research questions:

\begin{itemize}
	\item \textbf{RQ1: Is BugCerberus more accurate than existing bug localization approaches?} We assess the effectiveness of BugCerberus by comparing it against leading bug localization techniques using the SWE-bench-lite benchmark.
	
	\item \textbf{RQ2: How effectively does BugCerberus improve the existing automated issue fixing approaches?} We compare the issue-fixing performance of various approaches combined with our BugCerberus approach against those using baseline bug localization techniques, highlighting how BugCerberus improves overall fixing results.
	
	\item \textbf{RQ3: What are the contributions of key components to BugCerberus?} We conduct ablation studies on BugCerberus to evaluate the contribution of each key component of BugCerberus.
\end{itemize}

\subsection{Evaluation Dataset}

SWE-bench~\cite{Carlos2024} provides a dataset of 2,294 real-world software issues collected from 12 Python libraries. While SWE-bench-lite~\cite{Carlos2024} refines this dataset by filtering out low-quality issue descriptions, addressing these issues remains challenging due to the complexity of the underlying codebases. In line with most existing issue fixing work~\cite{chen2024coder, agentless}, we use SWE-bench and SWE-bench-lite as the datasets for our experiments. Specifically, the training and validation sets for the experiments are derived from the \textit{train} and \textit{dev} partitions of SWE-bench, while the \textit{test} set is sourced from the test partition of SWE-bench-lite. The training and validation sets were employed to fine-tune BugCerberus and the baselines, ensuring they did not overlap with the test set used for the evaluation.

\subsection{Baselines}
We compare BugCerberus against four bug localization baselines.

\begin{itemize}
	\item[-] RAG~\cite{jimenez2024swebench} is the representative automated issue fixing method designed to address issues in SWE-bench. In its bug localization component, RAG employs the traditional retrieval-based method BM25, referred to as \emph{RAG\_BL} in our experiments.
	
	\item[-] Agentless~\cite{agentless} is the state-of-the-art issue fixing method developed for SWE-bench, utilizing GPT-4o to recommend buggy elements at the file, function, and statement levels using specifically designed prompts. In our experiments, the bug localization component of Agentless is denoted as \emph{Agentless\_BL}.
	
	\item[-] FBL-BERT~\cite{Ciborowska2022} is the state-of-the-art bug localization approach based on LLM; this work uses BERT to locate bug, effectively aligning the semantics of bug report text with the inducing changeset.
	
	\item[-] LLMFL~\cite{Yang2024} is the latest language model-based fault localization method that identifies bugs without relying on test coverage information. To further assess the accuracy of BugCerberus in SWE-bench, we include a comparison with LLMFL, the latest fault localization technique.
\end{itemize}

\subsection{Metrics}
\textbf{Top-N}: Following the prior work~\cite{Meng2022, Yang2024, Zeng2022, Xu2024}, we compare Top-N to compare against state-of-the-art bug localization techniques. Top-N measures the number of bugs with at least one buggy element located within the first N positions (N=1, 3, 5, 10). 

\vspace{0.1cm}
\textbf{MAP}: Mean Average Precision (MAP) evaluates the average ranking position of all buggy elements identified by the bug localization method within the corresponding recommendation list. For instance, at the file level, MAP measures the ranked positions of buggy files in the file recommendation list. The formal definition is as follows:

\vspace{-0.5cm}
\begin{align*}
	\text{MAP} &= \frac{1}{n} \sum_{j=1}^{n} \text{AvgP}_j \\[-0.1pt]
	\text{AvgP}_j &= \frac{1}{|K_j|} \sum_{k \in K_j} \text{Prec@k} \\[-1pt]
	\text{Prec@k} &= \frac{1}{k} \sum_{i=1}^{k} \text{IsRelevant}(i)
\end{align*}
\noindent
where, $\text{AvgP}_j$ represents the average precision for the $j$-th bug, and $|K_j|$ denotes the total number of buggy elements associated with the $j$-th bug. $\text{Prec@k}$ indicates the precision of the top $k$ elements in the recommendation list, while $\text{IsRelevant}(i)$ returns 1 if the $i$-th element in the recommendation list is relevant to the bug and 0 otherwise. $K_j$ refers to the ranks of the buggy elements for the $j$-th bug.

\vspace{0.1cm}
\textbf{MRR}: Mean Reciprocal Rank (MRR) evaluates the position of the first buggy element identified by the bug localization method within the recommendation list. Its definition is as follows:

\vspace{-0.5cm}
\begin{align*}
	\text{MRR} = \frac{1}{n} \sum_{j=1}^n \frac{1}{\text{rank}_j}
\end{align*}

\noindent
where $rank_j$ represents the ranking position of the first buggy method modified to fix the $j$-th bug in the recommendation list.

\subsection{Implementation}

To learn buggy elements at the file, function, and statement levels, we implemented BugCerberus using three instances of Llama-3-8B-Instruct, a state-of-the-art large language model. We employed Llama Factory~\cite{zheng2024llamafactory}, an efficient fine-tuning framework for large models, to fine-tune the LLM with a default learning rate of 5e-5 and an effective batch size of 64. To ensure reproducibility by eliminating the influence of randomness, we set the model's generation parameters to a temperature of 0 and a top-p value of 1.0. The optimizer chosen is the commonly used Adam~\cite{Diederik2017}, which features an Adaptive Learning Rate to facilitate faster convergence. For learning rate tuning, we employ the Cosine Annealing scheduler~\cite{Ilya2017}, which adjusts the learning rate based on the cosine function.

We implement BugCerberus in Python using PyTorch. All results presented in our paper are obtained using an Intel(R) Xeon(R) Platinum 8352V CPU and a single A100-PCIE-40GB GPU installed with Ubuntu 20.04, CUDA 11.8. The time required for finetune of BugCerberus is about 30h.

\section{Experiment Results}

In this section, we provide detailed results and answers to each research question (RQ).

\begin{table*}[!t]
	\centering
	\caption{BugCerberus vs Other Bug Localization Techniques}
	\scalebox{1}{
		\rotatebox{0}{
			\begin{tabular}{llp{1.1cm}p{1.1cm}p{1.1cm}p{1.1cm}p{1.1cm}p{1.1cm}}
				\toprule
				\textbf{Level} & \textbf{Technique} & \multicolumn{6}{c}{\textbf{Metrics}} \\
				\cmidrule(lr){3-8}
				&                    & \textbf{Top1} & \textbf{Top3} & \textbf{Top5} & \textbf{Top10} & \textbf{MAP} & \textbf{MRR} \\
				\midrule
				\multirow{5}{*}{File} 
				& RAG\_BL~\cite{jimenez2024swebench} & 0.447 & 0.59 & 0.691 & 0.783 & 0.543 & 0.543 \\
				& Agentless\_BL~\cite{agentless} & 0.583 & 0.721 & \textbf{0.767} & \textbf{0.791} & 0.677&0.689 \\
				& FBL-BERT~\cite{Ciborowska2022}      & 0.549 & 0.619 & 0.656 & 0.702 & 0.424 & 0.455 \\
				& LLMFL ~\cite{Yang2024}       & 0.535 & 0.609 & 0.633 & 0.679& 0.399 & 0.421  \\
				& BugCerberus         & \textbf{0.651} & \textbf{0.745} & 0.754 & 0.763 & \textbf{0.694} & \textbf{0.733} \\
				\midrule
				\multirow{5}{*}{Function} 
				& RAG\_BL~\cite{jimenez2024swebench} & 0.046 & 0.051 & 0.051 & 0.066 & 0.043 & 0.043 \\
				& Agentless\_BL~\cite{agentless} & 0.356 & 0.483 & 0.502 & 0.507 & 0.286 & 0.287 \\
				& FBL-BERT~\cite{Ciborowska2022}      & 0.081 & 0.127 & 0.168 & 0.239 & 0.069 & 0.071 \\
				& LLMFL~\cite{Yang2024}        & 0.081 & 0.112 & 0.132 & 0.193 & 0.061 & 0.059 \\
				& BugCerberus         & \textbf{0.406} &\textbf{0.533}  &\textbf{0.569}  &\textbf{0.624}  &\textbf{0.425}  & \textbf{0.429} \\
				\midrule
				\multirow{5}{*}{Statement} 
				& RAG\_BL~\cite{jimenez2024swebench} & 0.009 & 0.014 & 0.014 & 0.014 & 0.007 & 0.007 \\
				& Agentless\_BL~\cite{agentless} & 0.230 & 0.354 & 0.364 & 0.368 & 0.210 & 0.211 \\
				& FBL-BERT~\cite{Ciborowska2022}     & 0.042 & 0.052 & 0.052 & 0.057 & 0.025 & 0.028 \\
				& LLMFL~\cite{Yang2024}         & 0.038 & 0.052 & 0.052 & 0.061 & 0.018 & 0.015 \\
				& BugCerberus         &\textbf{0.268}  & \textbf{0.373} & \textbf{0.401} & \textbf{0.453} & \textbf{0.326} & \textbf{0.332} \\
				\bottomrule
			\end{tabular}
	}}
	\label{comparision}
\end{table*}

\subsection{RQ 1. Overall Performance of BugCerberus}

\emph{1)} \textbf{Settings:} We evaluate the localization accuracy of BugCerberus against four baselines on the SWE-bench-lite dataset, using TopN, MAP, and MRR as evaluation metrics. Each localization method operates hierarchically, localizing bugs at the file, function, and statement levels. Since the Top 5 recommendations are the primary focus for most users~\cite{Yang2024, Xu2024}, the localizations at the function and statement levels depend on the Top 5 ranked results from the preceding level. The localization approaches that need to be fine-tuned, including FBL\_BERT, LLTAMO, and BugCerberus, are fine-tuned on the train set of the SWE-bench.

\begin{figure*}[!t]
	\centering
	\includegraphics[width=\textwidth]{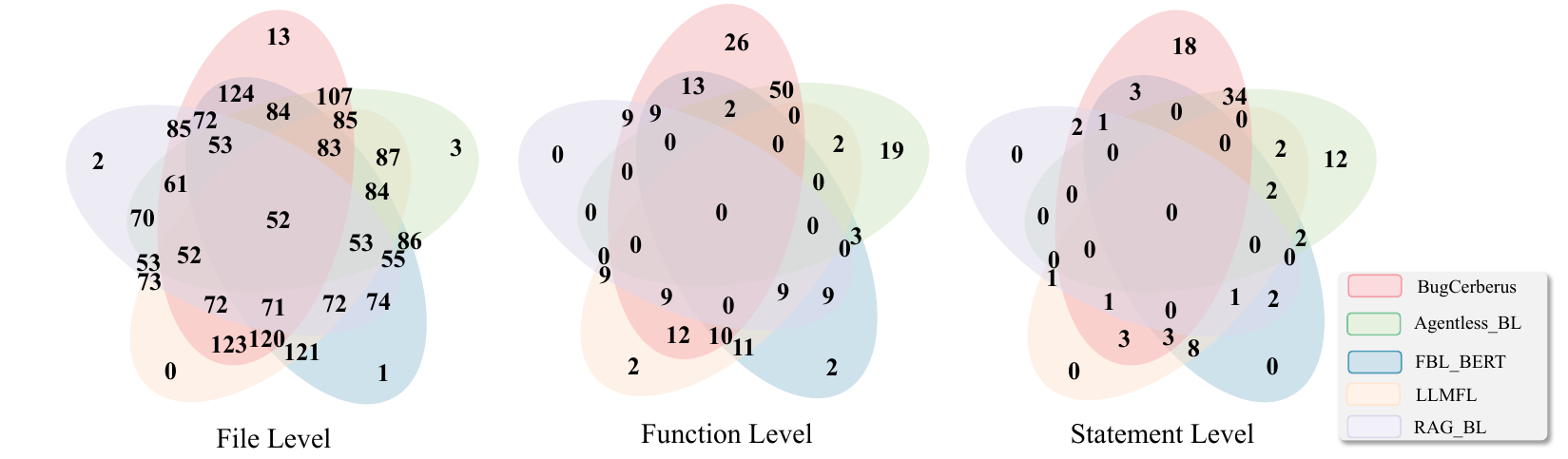}
	\caption{Overlap Analysis of BugCerberus and Other Bug Localization Techniques}
	\label{locCount}
	\vspace{-0.5cm}
\end{figure*}

\begin{figure*}[!t]
	\centering
	\includegraphics[width=\textwidth]{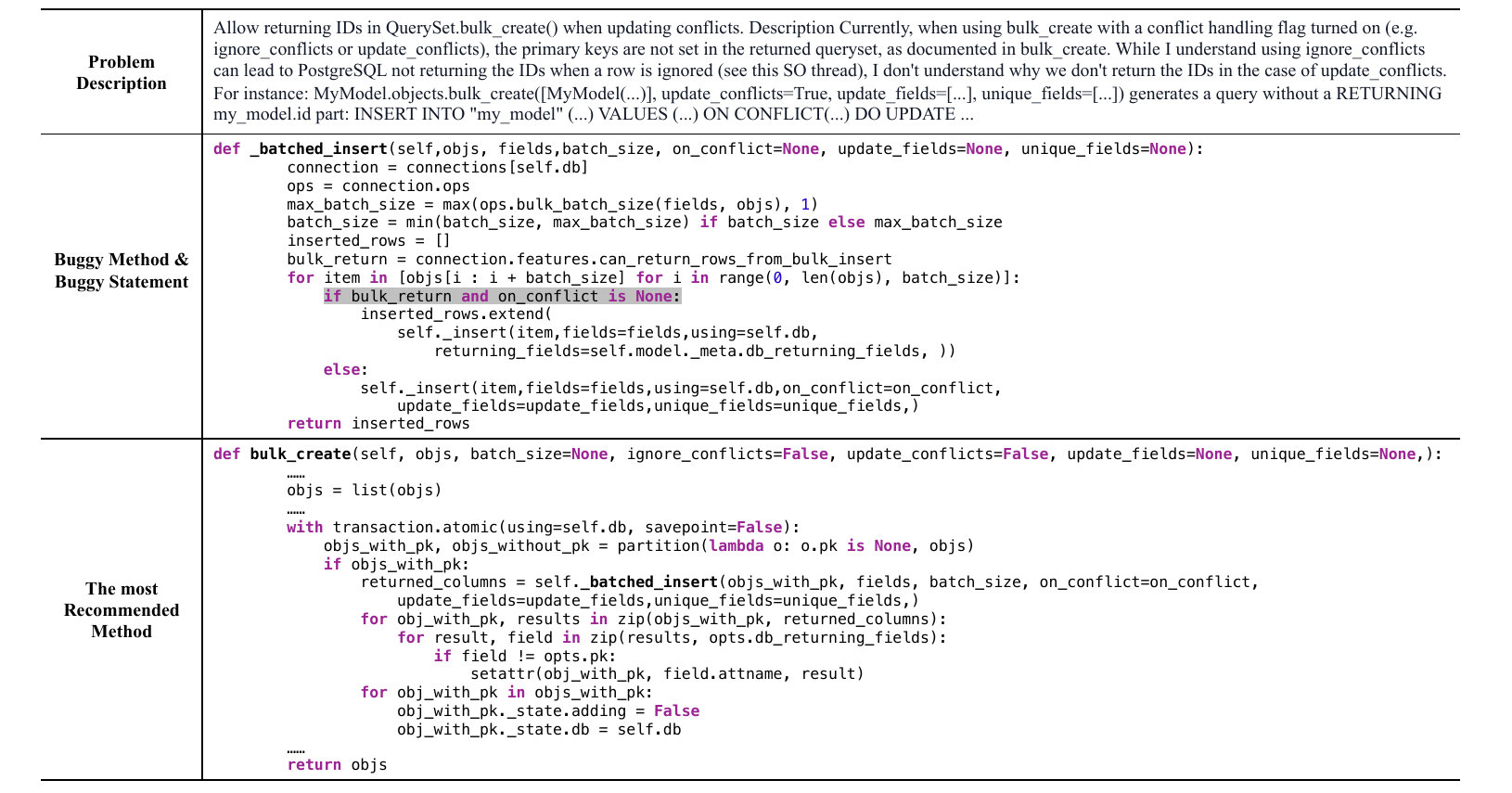}
	\caption{An Illustrative Example to Show How BugCerberus Locates the Bug Compared with the Baselines.}
	\label{expExampleForLoc}
	\vspace{-0.5cm}
\end{figure*}

\emph{2)} \textbf{Results:} The detailed results of the five bug localization techniques are presented in Table \ref{comparision}, which reports Top-N (N = 1, 3, 5, 10), MAP, and MRR for each technique. The bolded values in the table indicate the highest value in each column at each granularity level. The results demonstrate that BugCerberus outperforms all compared baselines at the function and statement levels and delivers competitive performance at the file level.

At the file level, BugCerberus achieves the highest values for Top-N (N = 1, 3), MAP, and MRR, with scores of 0.651, 0.745, 0.694, and 0.733, respectively, where MAP and MRR synthesize the effectiveness of localization. For Top 5 and Top 10, BugCerberus performs comparably to the top-performing Agentless\_BL. At the function level, BugCerberus surpasses the best baseline Agentless, improving Top-N (N = 1, 3, 5, 10), MAP, and MRR by 14.1\%, 10.4\%, 13.4\%, 23.1\%, 48.6\%, and 49.5\%, respectively. At the statement level, which offers the finest granularity, BugCerberus also achieves better performance, delivering improvements of 16.5\%, 5.4\%, 10.2\%,  23.1\%, 55.3\%, and 57.3\% for Top-N (N = 1, 3, 5, 10), MAP, and MRR, respectively, compared to the best baseline.

\emph{3)} \textbf{Analysis:} Notably, we observe that BugCerberus performs better with the finer granularity of localization, where function- and statement-level results outperform coarse-grained file-level results. In addition, we analyze the overlap of localization results within the Top 1 across the five approaches, as illustrated in Figure \ref{locCount}. At the file, function, and statement levels, BugCerberus identifies 13, 26, and 18 bugs, respectively, that baseline approaches fail to detect within the Top 1 results. Furthermore, the intersection analysis indicates that BugCerberus effectively captures the majority of bugs identified by other approaches, highlighting its accuracy across multiple granularity levels.

Furthermore, we analyzed the bugs that BugCerberus successfully located where four previous approaches failed. We found that BugCerberus exhibits a superior ability to understand the semantics of both the code and the problem description. For example, Figure \ref{expExampleForLoc} shows the problem description and part of the buggy method \emph{\_batched\_insert}, including the buggy line of code (highlighted) corresponding to the instance ID \emph{django\_django-17051}. The root cause of this issue is that the function \emph{bulk\_create} fails to return the primary key ID when handling a conflict. The problem description repeatedly emphasizes the failure of the function \emph{bulk\_create}, which led previous localization approaches to primarily recommend \emph{bulk\_create} as the source of the bug. However, the code snippet \emph{returned\_columns = self.\_batched\_insert} indicates that the return value of \emph{bulk\_create} (i.e., \emph{objs}) actually depends on\emph{ \_batched\_insert}. By leveraging the function call context and its semantics, the fine-tuned BugCerberus model can move beyond the problem description's emphasis on \emph{bulk\_create}, accurately identifying the buggy \emph{\_batched\_insert} method. This highlights BugCerberus's ability to analyze a deeper semantic association between the problem description and the underlying code.
\vspace{-0.5cm}
\begin{tcolorbox}[
	colback=gray!10, 
	colframe=black!70, 
	coltitle=black, 
	boxrule=0.75pt, 
	rounded corners, 
	drop shadow, 
	enhanced, 
	shadow={1mm}{-1mm}{0mm}{black!50}, 
	boxsep=0.1mm 
	]
	\textbf{Answer to RQ 1:} On the SWE-bench-lite dataset, BugCerberus surpasses all bug localization baselines, demonstrating its effectiveness in accurately identifying bugs. Specifically, compared to the best-performing baseline Agentless\_BL, BugCerberus achieves improvements of 12\%--17\%, 3\%--55\%, and 6\%--57\% in Top-1, MAP, and MRR across the three levels. In addition, BugCerberus outperforms Agentless\_BL by 14\%--17\%, 5\%--10\%, 10\%--13\%, 23\%, 48\%--55\%, and 50\%--57\% in Top-N (N = 1, 3, 5, 10), MAP, and MRR at the fine-grained function and statement levels.
\end{tcolorbox}

\subsection{RQ 2. Application of BugCerberus}

\begin{figure}[!t]
	\centering
	\includegraphics[width=0.5\textwidth]{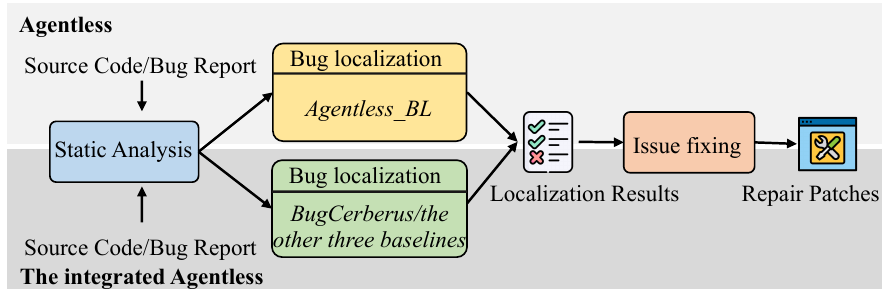}
	\vspace{-0.5cm}
	\caption{Constructing Pipeline of the Fixing Method Combined with BugCerberus}
	\label{aprsetting}
\end{figure}

\emph{1)} \textbf{Settings:} For this experiment, we selected the fixing method Agentless as it represents the latest approach and offers a fully open-source implementation. We integrated four bug localization approaches into Agentless, as illustrated in Figure \ref{aprsetting}, while Agentless\_BL is the bug localization approach used by Agentless. The integrated Agentless replaces the original bug localization part while retaining the same inputs. The subsequent program repair procedure utilizes the bug localization results produced by the five localization approaches. Following the Agentless patch generation process, 20 fix results were generated for the bugs in SWE-bench-lite, with each fix result containing 300 patches corresponding to 300 bugs. The optimal fix result was then selected from these 20 results to serve as the final fixing outcome. An issue is considered successfully fixed if the patch generated by Agentless passes all test cases. 

\begin{figure}[!t]
	\centering
	\includegraphics[width=0.5\textwidth]{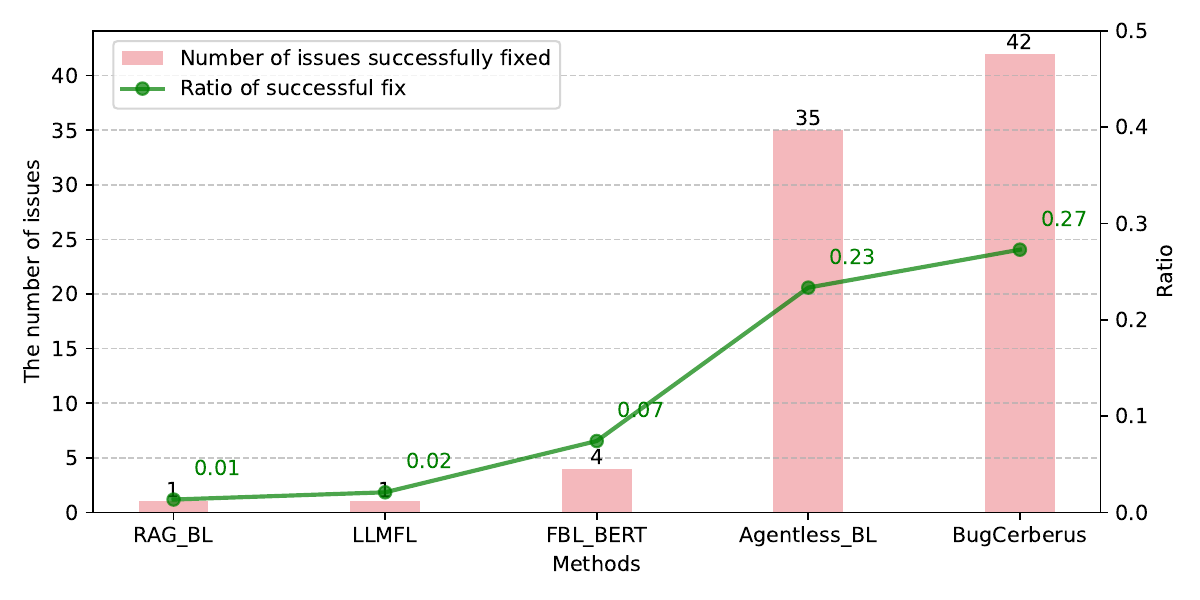}
	\vspace{-0.5cm}
	\caption{Issue Fixing Results with Five Bug Localization Techniques}
	\label{repairRatio}
\end{figure}

\emph{2)} \textbf{Results:}  The repair outcomes of the five approaches are presented in Figure \ref{repairRatio}. Each bar represents the number of issues successfully repaired by the fixing approach. The dashed line illustrates the ratio of successfully repaired issues to the number of issues with generated patches.

Figure \ref{repairRatio} illustrates that Agentless with BugCerberus generates the highest number of patches and successfully fixes 42 issues. In comparison, Agentless\_BL, FBL\_BERT, LLMFL, and SWE-Llama\_BL fix 35, 4, 1, and 1 issue, respectively. Compared to the best-performing baseline Agentless\_BL, BugCerberus resolves 7 more issues, achieving a 17.4\% improvement in the fix ratio.

\begin{figure*}[!t]
	\centering
	\includegraphics[width=\textwidth]{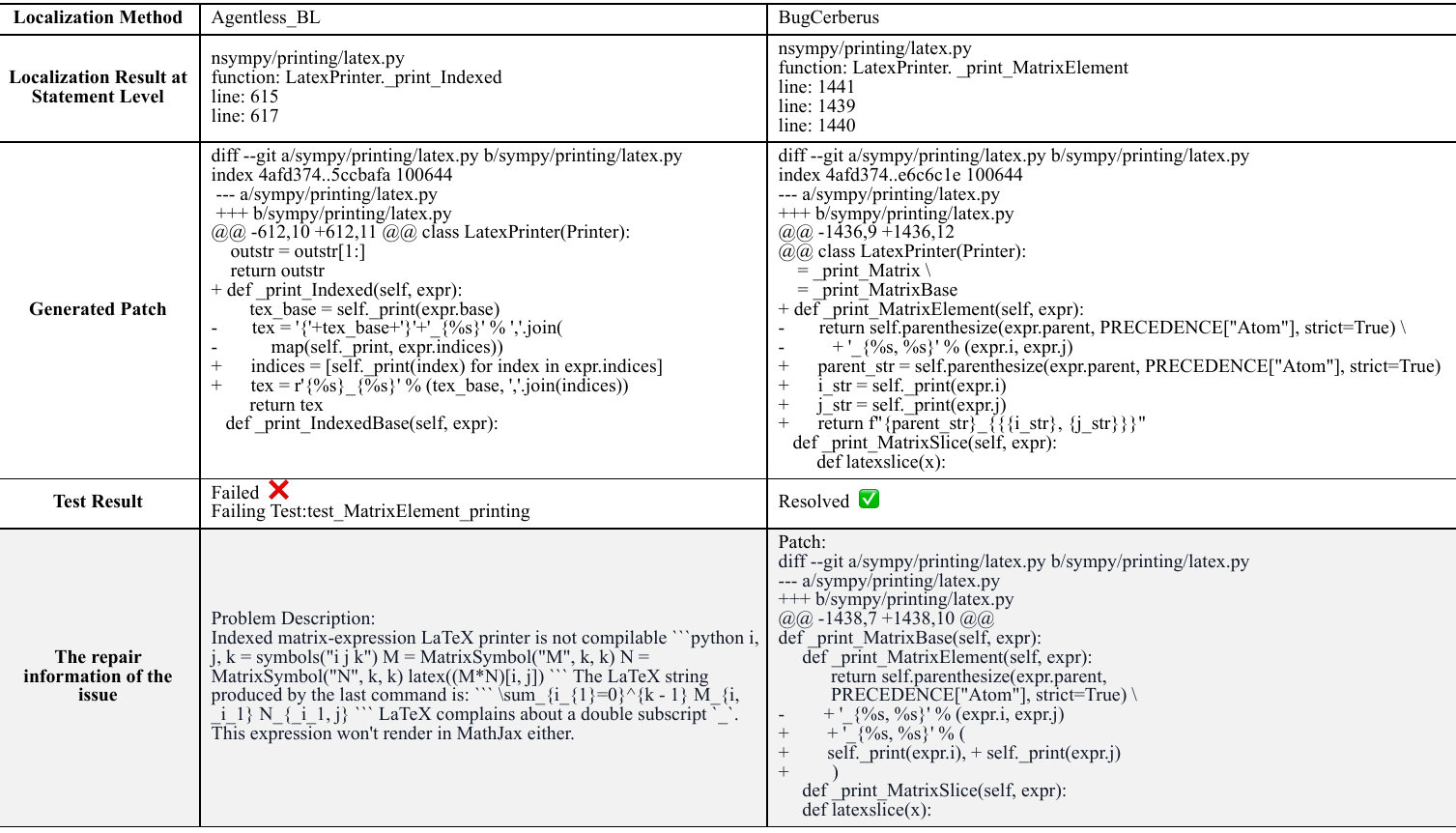}
	\caption{An Illustrative Example to Show How BugCerberus Helps Improve the Issue Fixing.}
	\label{repairExample}
	\vspace{-0.5cm}
\end{figure*}

\emph{3)} \textbf{Analysis:} From the experimental results, it is evident that BugCerberus, which exhibits the best localization results, also achieves the highest issue fixing result, with fixing rates of 27\%. Figure \ref{repairExample} presents a manual inspection of the successfully repaired issues by checking their instance IDs and the corresponding buggy lines in the generated patches. The result indicates that successful repairs are strongly linked to accurate localization results. Conversely, when the localization is incorrect, the repair process generates patches unrelated to the actual issue. Accurate localization is crucial for producing effective repair patches. For example, Figure \ref{repairExample} depicts the repair process for the issue with instance ID sympy\_\_sympy-15609. In this case, the buggy code is the \emph{+ '\_{\%s, \%s}' \% (expr.i, expr.j)} within the function\emph{ \_print\_MatrixElement}. Agentless\_BL identifies suspicious lines 615 and 617 in the unrelated function \emph{\_print\_Indexed}, leading to a patch targeting \emph{\_print\_Indexed}. This patch fails to address the buggy code and does not pass the test cases related to\emph{ \_print\_MatrixElement}. In contrast, Agentless with BugCerberus generates patches directly targeting the buggy lines in \emph{\_print\_MatrixElement}, successfully passing all related tests, albeit with patches differing from those provided in SWE-bench-lite.

\begin{tcolorbox}[
	colback=gray!10, 
	colframe=black!70, 
	coltitle=black, 
	boxrule=0.75pt, 
	rounded corners, 
	drop shadow, 
	enhanced, 
	shadow={1mm}{-1mm}{0mm}{black!50}, 
	boxsep=0.1mm 
	]
	\textbf{Answer to RQ 2:} Enhancing the bug localization phase significantly improves the overall issue fixing effectiveness. When Agentless is combined with BugCerberus, fixing result achieves a 17.4\% higher repair rate compared to the best-performing baseline.
\end{tcolorbox}

\subsection{RQ 3. Ablation Study}

\emph{1)} \textbf{Settings:} We investigated the contribution of key components to BugCerberus. The experiment focused on the extraction of static information within three levels and analyzed the impact of file-level and function-level localization on statement-level results. After removing the file-level localization, we recommend the Top 5 suspicious functions from the given repository and then locate statement-level bugs. After removing the function-level localization, we locate statement-level bugs from the recommended Top 5 files. To evaluate the contribution of these components, we utilized metrics such as Top-N, MAP, and MCC.

\begin{table*}[ht]
	\centering
	\caption{Ablation Study on Context Extraction}
	\scalebox{1}{
		\rotatebox{0}{
			\begin{tabular}{llcccccc}
				\toprule
				\textbf{Level} & \textbf{Static Information} & \multicolumn{6}{c}{\textbf{Metrics}} \\
				\cmidrule(lr){3-8}
				&                    & \textbf{Top1} & \textbf{Top3} & \textbf{Top5} & \textbf{Top10} & \textbf{MAP} & \textbf{MRR} \\
				\midrule
				\multirow{2}{*}{File} 
				& BugCerberus$_{no Structure}$         & 0.558 & 0.581 & 0.605 & 0.623 & 0.520 & 0.618 \\
				& BugCerberus         & 0.651 & 0.745 & 0.754 & 0.763 & 0.694 & 0.733 \\
				& Improvement (\%)         & $\uparrow$16.7\% & $\uparrow$28.2\% & $\uparrow$24.7\% & $\uparrow$22.5\% & $\uparrow$33.5\% & $\uparrow$18.6\% \\
				\midrule
				\multirow{2}{*}{Function} 
				& BugCerberus$_{no Call Chain}$         & 0.269 & 0.355 & 0.426 & 0.508 & 0.281 & 0.286 \\
				& BugCerberus         & 0.406 & 0.533 & 0.569 & 0.624 & 0.425 & 0.429 \\
				& Improvement (\%)        & $\uparrow$50.9\% & $\uparrow$50.1\% & $\uparrow$33.6\% & $\uparrow$22.8\% & $\uparrow$51.2\% & $\uparrow$50.0\% \\
				\midrule
				\multirow{2}{*}{Statement} 
				& BugCerberus$_{no Program Slice}$         & 0.202 & 0.301 & 0.344 & 0.425 & 0.301 & 0.319 \\
				& BugCerberus         & 0.268 & 0.373 & 0.401 & 0.453 & 0.326 & 0.332 \\
				& Improvement (\%)          & $\uparrow$32.7\% & $\uparrow$23.9\% & $\uparrow$16.6\% & $\uparrow$6.6\% & $\uparrow$8.3\% & $\uparrow$4.1\% \\
				\bottomrule
			\end{tabular}
	}}
	\label{ablation_component}
\end{table*}

\begin{table}[ht]
	\centering
	\caption{Ablation Study on Bug Localization Components at File and Function Level}
	\scalebox{0.85}{
		\rotatebox{0}{
			\begin{tabular}{llccccc}
				\toprule
				\textbf{Hierarchical Component}& \multicolumn{6}{c}{\textbf{Metrics}} \\
				\cmidrule(lr){2-7}
				& \textbf{Top1} & \textbf{Top3} & \textbf{Top5} & \textbf{Top10} & \textbf{MAP} & \textbf{MRR} \\
				\midrule
				No File Level   & 0 & 0 & 0.023 & 0.042 & 0.013 & 0.022 \\
				\midrule
				No Function Level        & 0.092& 0.183 & 0.275 & 0.35 & 0.125 & 0.132 \\
				\midrule
				BugCerberus         & 0.268 & 0.373 & 0.401 & 0.453 & 0.326 & 0.332 \\
				\bottomrule
			\end{tabular}
	}}
	\label{ablation_level}
\end{table}

\emph{2)} \textbf{Effect of extracting static information:} Table \ref{ablation_component} presents the contributions of the three levels of static analysis components. In Table \ref{ablation_component}, the first row for each level shows the localization results after removing the corresponding component, while the third row illustrates the improvement in BugCerberus when the component is added. According to the data in Table 2, BugCerberus achieves better results compared to its performance after removing the three components at different levels, indicating that static information analysis and context extraction at each layer are effective for BugCerberus. Furthermore, the contribution of static information in each layer is significant, particularly at the function level, where Top-N (N = 1, 3, 5) improved by 50.9\%, 50.1\%, and 33.6\%, respectively. The synthesized evaluation metrics MAP and MRR increased by 51.2\% and 50.0\%. For the context components at the file and statement levels, the Top 5 improved by 24.7\% and 16.6\%, while MAP improved by 33.5\% and 8.3\%, respectively. These results demonstrate that static context extraction is a crucial component of BugCerberus, significantly enhancing the model’s ability to learn bug-related features.

\emph{3)} \textbf{Effect of hierarchical localization:} Table \ref{ablation_level} highlights the contributions of file-level and statement-level localization components. According to the data presented in Table \ref{ablation_level}, the absence of either the file-localization component or the function-localization component significantly compromises the effectiveness of statement-level bug localization. Specifically, when the file-level localization component is removed, BugCerberus's statement-level localization performance exhibits a 94.3\% reduction in the Top 5 compared to the original result and a 96\% reduction in MAP. Similarly, removing the function-level localization component leads to a 31.4\% decrease in Top 5 accuracy and a 61.6\% decrease in MAP. This reduction can be attributed to the ability of file-level and function-level localization to effectively narrow the search space for identifying suspicious code. On average, projects in the SWE-bench dataset contain 11020 functions and 168721 code statements, with one file containing 152 statements and one function containing 15 statements. Locating a bug within 15 lines of a function is considerably more efficient than searching across 168,721 lines of an entire project. Therefore, The results highlight the critical role of the file-localization and function-localization components in BugCerberus. Both components enhance accuracy and overall effectiveness in identifying bugs by significantly reducing the scope of bug localization.

\begin{tcolorbox}[
	colback=gray!10, 
	colframe=black!70, 
	coltitle=black, 
	boxrule=0.75pt, 
	rounded corners, 
	drop shadow, 
	enhanced, 
	shadow={1mm}{-1mm}{0mm}{black!50}, 
	boxsep=0.1mm 
	]
	\textbf{Answer to RQ 3:} This experiment demonstrates the effectiveness of context extraction components and hierarchical localization components in enhancing the localization performance of BugCerberus. Notably, the absence of the file-level hierarchical localization component results in a significant 94.3\% reduction in the Top 5 accuracy of BugCerberus.
\end{tcolorbox}

\section{Threats to Validity}

\textbf{Internal Validity} concerns the potential risks of data leakage and the annotation of buggy elements. Since SWE-bench's data is collected from the open-source platform GitHub, the same source used for training the original Llama model, there is a risk of data leakage. However, as demonstrated in Section 5.3, BugCerberus performs poorly without context extraction and hierarchical localization components, suggesting that the impact of potential data leakage is limited. Another internal threat lies in the annotation of buggy elements, as the criteria for determining whether a code element is buggy may be different. To mitigate this threat, we adopt the widely-used standard established in previous works~\cite{Meng2022, Spencer2017}, which labels deleted or modified code as buggy.

\textbf{External Validity} pertains to the generalizability of the results. BugCerberus is currently evaluated only on Python projects, as it is specifically designed to locate issues in SWE-bench and enhance SWE-bench-based automated issue fixing techniques. Both the dataset and repair approaches are tailored to Python. Nevertheless, the contextualized instruction construction and static information processing techniques employed by BugCerberus can be extended to other programming languages with appropriate adaptations.

\section{Related Work}

This section provides an overview of related work in bug localization and existing approaches in GitHub issue fixing.

\subsection{Bug localization}

Bug localization involves identifying the relevance between a bug report and program elements, typically by recommending the most relevant program components~\cite{Yang2021}. Earlier approaches in bug localization extensively employed the Support Vector Model (SVM) due to their proven effectiveness. For instance, Zhou et al.~\cite{Zhou2012} introduced BugLocator, which utilizes a revised SVM to calculate the similarity between a bug report and source code. It then recommends Top-N suspicious code files by combining this similarity with historical code evolution data. As deep learning architectures demonstrated their ability to capture contextual information in natural language processing, subsequent studies began leveraging these techniques for bug localization. For example, Kim et al.~\cite{Kim2013} proposed a two-phase model that predicts suspicious files based on the content and quality of bug reports. More recently, the emergence of a large language model (LLM) with advanced code comprehension capabilities has further advanced bug localization. For example, Ciborowska et al.~\cite{Ciborowska2022} introduced FBL-BERT, the first application of BERT, to calculate the correlation between change code and bug reports.

While these approaches are effective, most research has focused on Java and C programming languages, with limited attention to Python~\cite{Mohammad2024}. Additionally, many existing approaches concentrate on a single granularity of localization, such as file-level or function-level recommendations, which do not fully address the multi-level (file, function, statement) needs of users. Unlike existing bug localization techniques, our work, BugCerberus, focuses on Python and aims to learn across multiple levels of granularity. We utilize specialized models tailored to each granularity, ensuring accurate and comprehensive bug localization to meet the unique requirements at the file, function, and statement levels.

\subsection{GitHub Issue Fixing}
GitHub issue fixing is tasked to repair the GitHub issue using the given issue information automatically. Several approaches have been proposed for this task. For instance, RAG~\cite{jimenez2024swebench} employs the BM25 retriever to search for bug-related code files and trains the Llama model to address the issue based on the retrieved files. 
CODER~\cite{chen2024coder} integrates BM25 and Ochiai retrievers to identify suspicious files, which are then used as inputs for the repair agent. 
SWE-Agent~\cite{yang2024sweagent} leverages GPT-4o to consult relevant files and code contents using commands such as find\_file, search\_file, and search\_dir. 
Lastly, Agentless~\cite{agentless} queries GPT-4o for relevant code elements at various levels (file, function, and statement) based on a problem description and designed prompts.

Bug localization is essential in GitHub issue fixing. However, most bug localization techniques in issue-fixing approaches rely on coarse-grained file-level similarity computations. Even those with fine-grained localization often overlook contextual inconsistencies across different levels (file, function, and statement) and fail to provide comprehensive code characterization at each granularity level.

In contrast, our proposed BugCerberus addresses these gaps by focusing on code feature extraction and learning across multiple levels of granularity. BugCerberus fine-tunes three specialized models to meet the unique localization requirements at each granularity, ensuring accurate and comprehensive bug localization for Python programs.

\section{Conclusion and Future Work}

In this paper, we present BugCerberus, a novel bug localization framework designed to identify buggy elements across three levels: file, function, and statement. The framework begins by leveraging static program analysis to extract contextual information for buggy elements at each level. Subsequently, three LLMs are instruction fine-tuned to effectively learn the unique features of these elements. Using the fine-tuned models, BugCerberus performs hierarchical bug localization, progressing from files to functions and finally to statements. Furthermore, we integrate BugCerberus into an automated issue fixing method to address SWE-bench-lite issues, enabling an investigation into the impact of the bug localization phase on issue fixing. Experimental results demonstrate that BugCerberus surpasses existing bug localization approaches in identifying buggy code within SWE-bench-lite and enhances the repair performance of the issue fixing method Agentless by 17.4\%. 

Our work is the first exploration of bug localization within the context of automated issue fixing. In the future, we aim to categorize the issues in the SWE-bench and develop tailored bug localization approaches for each specific classification.

\bibliographystyle{IEEEtran}
\bibliography{reference}

\vspace{-0.8cm}
\begin{IEEEbiography}[{\includegraphics[width=1in,height=1.25in,clip,keepaspectratio]{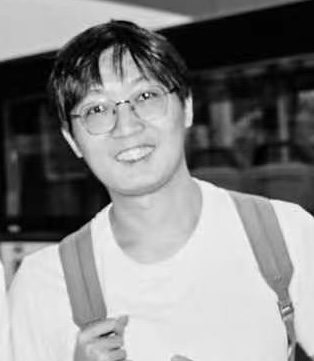}}]{Jianming Chang}
 is working toward a Ph.D. degree at the Southeast University, Nanjing, China. He also serves as a research assistant at Singapore Management University. His research interests include program analysis and program slicing. He currently focuses on bug localization and automated issue fixing. 
\end{IEEEbiography}
\vspace{-0.5cm}

\begin{IEEEbiography}[{\includegraphics[width=1in,height=1.25in,clip,keepaspectratio]{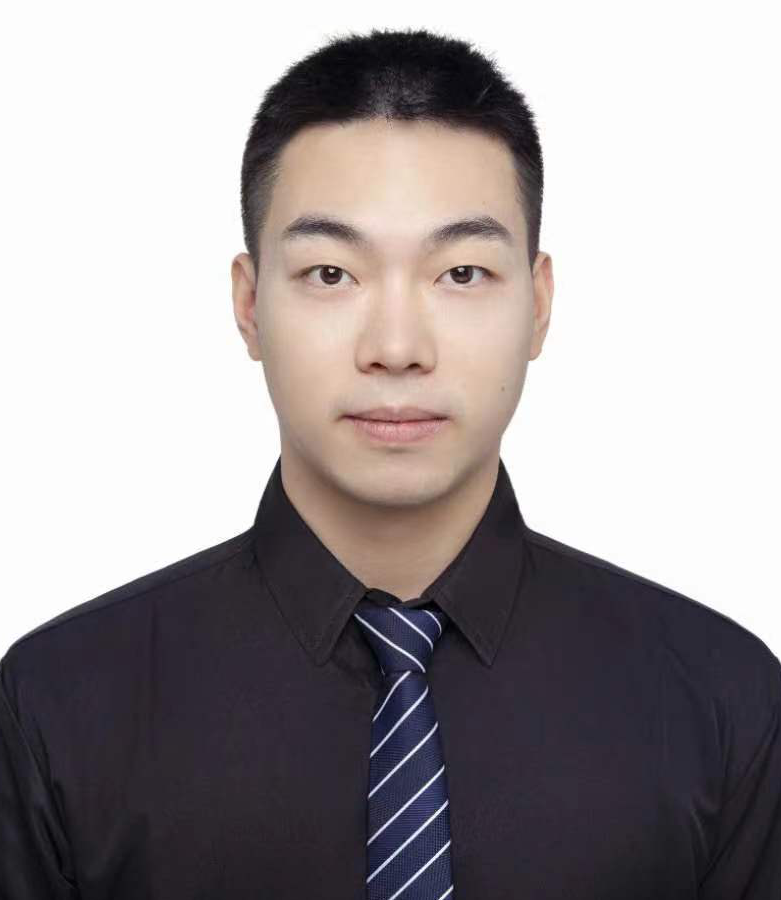}}]{Xin Zhou}
received the Ph.D. degree in computer science from Singapore Management University, Singapore. He is a Postdoctoral researcher with Singapore Management University. His research interest is in code representation and software activity automation, such as maintenance, debugging, and code reviewing. He is currently focusing on designing novel pretrained code representations and applying them to automate developer-intensive activities.
\end{IEEEbiography}
\vspace{-0.5cm}
\begin{IEEEbiography}[{\includegraphics[width=1in,height=1.25in,clip,keepaspectratio]{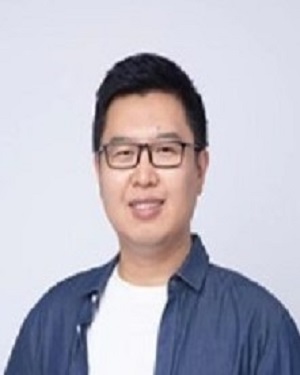}}]{Lulu Wang}
	is an associate professor of Computer Science and Engineering School at the Southeast University, Nanjing, China. His re- search interests include: path profiling, pro- gram analysis and program slicing. He got his Bachelor‘s degree in Computer Science from Southeast University (China) in 2006, and his Doctor‘s degree in Software Engineering from Southeast University (China) in 2012.
\end{IEEEbiography}
\vspace{-0.5cm}
\begin{IEEEbiography}[{\includegraphics[width=1in,height=1.25in,clip,keepaspectratio]{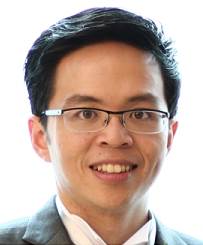}}]{David Lo}
(Fellow, IEEE) is a Professor in computer science and the Director of the information and systems cluster at the School of Computing and Information Systems, Singapore Management University. He leads the Software Analytics Research (SOAR) group. His research interest is in the intersection of software engineering, cybersecurity, and data science, encompassing socio-technical aspects and analysis of different kinds of software artifacts, with the goal of improving software quality and security and developer productivity.
\end{IEEEbiography}
\vspace{-0.5cm}
\begin{IEEEbiography}[{\includegraphics[width=1in,height=1.25in,clip,keepaspectratio]{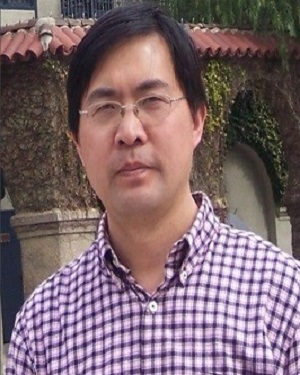}}]{Bixin Li}
	is a professor of Computer Science and Engineering School at the Southeast Uni- versity, Nanjing, China. His research inter- ests include: Program slicing and its appli- cation; Software evolution and maintenance; and Software modeling, analysis, testing and verification. He has published over 90 articles in refereed conferences and journals. He leads a Software Engineering Institute in Southeast University, and over 20 young men and women are hard working on national and international projects.
	
\end{IEEEbiography}

\vfill

\end{document}